\documentclass[12pt, reqno]{amsart}
\usepackage{geometry}

\usepackage[utf8]{inputenc}
\usepackage[all, cmtip]{xy}

\usepackage{macros}
\renewcommand{\O}{\mathcal{O}}
\renewcommand{\P}{\mathcal{P}}
\DeclareMathOperator{\Op}{Op}

\newcommand{\C}{\cat{C}}
\newcommand{\D}{\cat{D}}
\newcommand{\pos}{\real_{> 0}}
\newcommand{\nonneg}{\real_{\geq 0}}
\newcommand{\extreal}{\overline{\real}}
\newcommand{\convrelC}{\cat{ConvRel}}

\newcommand{\Ssh}{S_{\mathrm{sh}}} 
\newcommand{\Sbath}{S_{\mathrm{bath}}} 
\newcommand{\Sgas}{S_{\mathrm{ideal}}} 
\newcommand{\Stank}{S_{\mathrm{tank}}} 

\title{Compositional Thermostatics}

\author[Baez]{John C.\ Baez$^{1,2}$}
\author[Lynch]{Owen Lynch$^{3,4}$}
\author[Moeller]{Joe Moeller$^{5}$}

\address{$^1$Department of Mathematics, University of California, Riverside, CA 92521, USA}
\address{$^2$Centre for Quantum Technologies, National University of Singapore, Singapore 117543}
\address{$^3$Mathematics Department, Universiteit Utrecht, Heidelberglaan 8, 3584 CS Utrecht, The Netherlands}
\address{$^4$Topos Institute, 2140 Shattuck Ave \#610, Berkeley, CA 94704, USA}
\address{$^5$National Institute of Standards and Technology, 100 Bureau Dr, Gaithersburg, MD 20899, USA \vskip 1em}

\email{baez@ucr.edu, owen@topos.institute, joseph.moeller@nist.gov}

\begin{document}


\begin{abstract}
We define a thermostatic system to be a convex space of states together with a concave function sending each state to its entropy, which is an extended real number. This definition applies to classical thermodynamics, classical statistical mechanics, quantum statistical mechanics, and also generalized probabilistic theories of the sort studied in quantum foundations. It also allows us to treat a heat bath as a thermostatic system on an equal footing with any other. We construct an operad whose operations are convex relations from a product of convex spaces to a single convex space, and prove that thermostatic systems are algebras of this operad. This gives a general, rigorous formalism for combining thermostatic systems, which captures the fact that such systems maximize entropy subject to whatever constraints are imposed upon them. 
\end{abstract}

\maketitle

\section{Introduction}

A large part of thermodynamics deals with systems in equilibrium: this deserves to be called `thermostatics'. To treat this subject in a modern mathematical spirit, we define a thermostatic system to be any convex space of `states' together with a concave function assigning to each state its entropy. Whenever several such systems are combined and allowed to come to equilibrium, the new equilibrium state maximizes the total entropy subject to constraints. We explain how to express this idea in a rigorous and fully general way using an operad. Intuitively speaking, the operad we construct has as operations \emph{all possible ways of combining thermostatic systems}. For example, there is an operation that combines two gases in such a way that they can exchange energy and volume, but not particles---and another operation that lets them exchange only particles, and so on.

Operads provide a way to take the business of combining physical systems, often left to informal rules of thumb, and turn it into mathematics.  Not only is this a prerequisite for proving general theorems about compositionality, it can also serve as the basis for software.  For example, the AlgebraicJulia project has produced several software packages based on operads that let users build complex models of dynamical systems by composing simpler parts.  The AlgebraicPetri package does this with Petri nets \cite{BFLP}, one common framework for describing systems of ordinary differential equations, while the StockFlow package does it using a related framework: stock and flow diagrams \cite{BLLOP}.  The Decapodes package uses operads to help users build `multiphysics models' involving partial differential equations for electromagnetism, fluid mechanics and the like \cite{BFHP}.   It is natural to extend this methodology to handle systems of other sorts, including thermostatic systems.  But first the underlying mathematics must be worked out.

Our approach requires an abstract kind of convex space that need not be a subset of a vector space, described in \cref{sec:convex_spaces}. In \cref{sec:thermostatic_systems} we show how this lets us systematically handle thermostatic systems in many contexts, including classical thermodynamics, classical statistical mechanics, and quantum statistical mechanics. Even the `heat bath' becomes a rigorously well-defined thermostatic system on an equal footing with the rest. In \cref{sec:entropy_maximization} we study the entropy maximization principle for general thermostatic systems, and in \cref{sec:operads} we use this to describe compositional thermostatics using an operad. We end with a variety of examples.

Starting perhaps with the work of Gudder \cite{Gudder1977}, abstract convex spaces have also become important in the foundations of quantum mechanics, where they are used to study both states and effects in so-called `generalized probabilistic theories' \cite{Jacobs}. Entropy has been studied in the context of these generalized probabilistic theories \cite{Barnum_et_al, Krumm_et_al,ShortWehner}, and in \cref{ex:generalized} we show our framework applies also to these.

\subsubsection*{Acknowledgements}

We thank Spencer Breiner, Tobias Fritz, Tom Leinster and Sophie Libkind for helpful discussions. We thank the Topos Institute for supporting this research.

\section{Convex spaces}
\label{sec:convex_spaces}

The central object in our thermostatics formalism is a notion of `convex spaces' that need not be convex subsets of a vector space. 

\begin{definition}
    A \define{convex space} is a set $X$ with an operation $c_\lambda \colon X \times X \to X$ for each $\lambda \in [0, 1]$ such that the following identities hold:
    \begin{itemize}
        \item $c_1(x, y) = x$, 
        \item $c_\lambda(x, x) = x$, 
        \item $c_\lambda(x, y) = c_{1-\lambda}(y, x)$, 
        \item $c_\lambda(c_\mu(x, y) , z) = c_{\lambda'}(x, c_{\mu'}(y, z))$ for all $0 \le \lambda, \mu, \lambda', \mu' \le 1$ satisfying $\lambda\mu = \lambda'$ and $ 1-\lambda = (1-\lambda')(1-\mu')$.
    \end{itemize}
Given a set $X$, a \define{convex structure} on $X$ is a collection of functions $c_\lambda \maps X \times X \to X$ for $\lambda \in [0,1]$ obeying the above axioms.
\end{definition}

\begin{example}
Any vector space is a convex space with the convex structure $c_\lambda(x, y) = \lambda x + (1-\lambda)y$.
\end{example} 

The abstract definition of a convex space has been reinvented many times \cite{Fritz}, but perhaps the story starts in 1949 with Stone's `barycentric algebras' \cite{Stone}. Beside the above axioms, Stone included a \define{cancellation axiom}: whenever $\lambda \ne 0$, 
\[  c_\lambda(x,y) = c_\lambda(x',y) \implies x = x' . \]
This allowed him to prove that any barycentric algebra is isomorphic to a convex subset of a vector space. Later Neumann \cite{Neumann} noted that a convex space, defined as above, is isomorphic to a convex subset of a vector space if and only if the cancellation axiom holds. 

Dropping the cancellation axiom has convenient formal consequences, since the resulting more general convex spaces can then be defined as algebras of a finitary commutative monad \cite{Jacobs,Swirszcz}, giving the category of convex spaces very good properties. But dropping this axiom is no mere formal nicety. We need the set of possible values of entropy to be a convex space. One candidate is the set $\nonneg = [0,\infty)$. However, for a well-behaved formalism based on entropy maximization, we want the supremum of \emph{any} set of entropies to be well-defined. This forces us to consider the larger set $[0,\infty]$, which does not obey the cancellation axiom. But in fact, our treatment of the heat bath starting in \cref{ex:heat_bath} forces us to consider \emph{negative} entropies---not because the heat bath can have negative entropy, but because the heat bath acts as an infinite reservoir of entropy, and the change in entropy from its default state can be positive or negative. This suggests letting entropies take values in the convex space $\real$, but then the requirement that any set of entropies have a supremum (including empty and unbounded sets) forces us to use the larger convex space $\extreal = [-\infty,\infty]$, which does not obey the cancellation axiom.

Of course, convexity has been widely used already in classical thermodynamics, in particular for studying the Legendre transform \cite{GaglianiScotti, PointErlicher, Willerton}. Based on this and other applications, convex analysis has grown into quite a large subject: see Rockafellar's book \cite{Rockafellar}. This will become important in future developments, but note that his text only considers convex subsets of $\real^n$. 

We now consider some more examples of convex spaces:

\begin{definition}
    A subset $S$ of a convex space $X$ is a \define{convex subspace} if for all $x, x' \in S$ and all $0 \le \lambda \le 1$ we have $c_\lambda(x,x') \in S$.
\end{definition}

A convex subspace of a convex space is a convex space in its own right. 

\begin{example}
    The \define{positive orthant} $\pos^n \ins \real^n$ is the subset of $\real^n$ consisting of vectors with all positive coordinates: $x \in \real^n$ such that $x_i > 0$ for all $i$. This is a convex subspace of $\real^n$, and thus a convex space in its own right.
\end{example}

\begin{example}
\label{ex:simplex}
    The \define{$n$-simplex} $\Delta^n \ins \real^{n+1}$ is the set of probability distributions on the set $\{0,\dots, n\}$:
    \[\Delta^n = \set{x \in \real^{n+1} \st x_i \geq 0, \sum_{i=0}^n x_i = 1} .\]
    This is a convex subspace of the vector space $\real^{n+1}$, and thus a convex space in its own right.
\end{example}

The next example does not obey the cancellation axiom:

\begin{example} \label{ex:extended_reals}
    The set $\extreal$ of extended reals has a unique convex structure with
    \[ 
    \begin{array}{rclll}
        c_\lambda(x,y) &=& \lambda x + (1-\lambda) y &&\text{for $x,y \in \real$} \\ c_\lambda(x,\infty)
        &=& \; \infty &&\text{for $x \in \real$} \\
        c_\lambda(x,-\infty) &=& -\infty &&\text{for $x \in \real$} \\
        c_\lambda(-\infty, \infty) &=& -\infty 
    \end{array}
    \]
    for all $\lambda \in (0,1)$. This convex structure extends the usual one on $\real$. To see that these operations indeed obey the laws of a convex structure, note that if $X$ is any convex space and $\{\ast\}$ is some singleton, there is a unique convex structure on the disjoint union $X \sqcup \{\ast\}$ extending that on $X$ such that $c_\lambda(x,\ast) = \ast$ for all $\lambda \in (0,1)$. Using this trick once, we get a convex structure on $(-\infty,\infty] = \real \sqcup \{\infty\}$. Using it again, we get the desired convex structure on $\extreal = (-\infty,\infty] \sqcup \{-\infty\}$. Note the asymmetry: whenever we take a nontrivial convex combination of $\infty$ and $-\infty$, we get $-\infty$. There is another convex structure on $\extreal$ with $c_\lambda(-\infty, \infty) = \infty$ for $\lambda \in (0,1)$. However, our choice is physically motivated: with the other choice, Lemma \ref{thm:functor} would not hold.
\end{example}

We will now consider several notions of maps between convex spaces. The first notion is perhaps the most straightforward: a function that preserves convex combinations.

\begin{definition}
    A \define{convex-linear map} from a convex space $X$ to a convex space $Y$ is a convex relation $f \subseteq X \times Y$ that is a function.
    Equivalently, a convex-linear map is a function $f \maps X \to Y$ such that for $x, x' \in X$ and all $\lambda \in [0,1]$, 
    \[   f(c_\lambda(x,x')) = c_\lambda(f(x), f(x')).\]
\end{definition}

\begin{example}
    If $V$ and $W$ are vector spaces, any linear map $L \maps V \to W$ is convex-linear. Any affine map is also convex-linear.
\end{example}

One extension of convex maps can be given in the case that we are mapping into a convex space with an ordering: we then can relax the equality of convex-linearity to an inequality.

\begin{definition} 
    Given a convex space $X$, a function $f \maps X \to \extreal$ is \define{concave} if for all $x, x' \in X$ and all $0 \le \lambda \le 1$, 
    \[   f(c_\lambda(x, x')) \geq c_\lambda(f(x), f(x'))\]
\end{definition}

\begin{example}
    $\log \maps \pos \to \extreal$ is concave.
\end{example}

Note that if we had chosen the convex structure on $\extreal$ with $c_\lambda(\infty, -\infty) = \infty$ for $\lambda \in (0, 1)$ then a concave function $f \maps X \to \extreal$ with $f(x) = \infty$ and $f(x') = -\infty$ would need to be infinite on all nontrivial convex linear combinations of $x$ and $x'$, since
\[ f(c_\lambda(x, x')) \geq c_\lambda(f(x), f(x')) = \infty. \]
With the convex structure we actually chose for $\extreal$, we merely need
\[ f(c_\lambda(x, x')) \geq c_\lambda(f(x), f(x')) = -\infty, \]
which is automatic. In fact, we \emph{need} this looser second requirement in the proof of Lemma \ref{thm:functor}, which is crucial to our work.

Just as we can generalize functions between sets to relations between sets, we can also generalize convex-linear maps to \emph{convex relations}. To define this, we first define the product of two convex spaces.

\begin{definition} \label{def:product_convex_spaces}
    Given two convex spaces $X$ and $Y$, we may form their product, $X \times Y$. This has a convex structure given by
    \[ c_\lambda((x, y), (x', y')) = (c_\lambda(x, x'), c_\lambda(y, y'))\]
\end{definition}

\begin{definition}
    A \define{convex relation} from a convex space $X$ to a convex space $Y$ is a convex subspace of $X \times Y$.
\end{definition}

\begin{example} \label{ex:graph_of_convex-linear}
    If $f \maps X \to Y$ is any convex-linear map, then its graph \[\set{ (x,y) \in X \times Y \mid y = f(x) } \] is a convex relation.
\end{example}

\begin{example}
    If $f \maps X \to \extreal$ is any concave map, then its subgraph \[\set{ (x, y) \in X \times Y \mid y \leq f(x) }\] is a convex relation.
\end{example}

We will often think of convex relations in terms of ``compatibility.'' That is, a convex relation $R \subseteq X \times Y$ expresses when some description of the system $x \in X$ is ``compatible'' with another description $y$. This compatibility need not necessarily be functional: there could be any number of descriptions $y$ compatible with $x$. Thus, we use relations.  We shall see many examples of convex relations in Section
\ref{sec:entropy_maximization}

\begin{definition} \label{def:composition_of_convex_relations}
    Composition of convex relations is defined in the following way. If $R \ins X \times Y$ and $R' \ins Y \times Z$, we define their composite $R' \circ R \ins X \times Z$ by
    \[ R' \circ R = \set{(x,z) \in X \times Z \st \exists y \in Y, (x,y) \in R, (y,z) \in R'} \]
\end{definition}

\begin{proposition} 
   The composition of two convex relations is convex.
\end{proposition}

\begin{definition}
    Let $\convrelC$ denote the category of convex space and convex relations, with composition as defined in \cref{def:composition_of_convex_relations}. Let $\convC$ denote the subcategory of convex spaces and convex-linear maps, where the inclusion is given by graphs as in \cref{ex:graph_of_convex-linear}.
\end{definition}

\section{Thermostatic systems}
\label{sec:thermostatic_systems}

A thermostatic system is a convex space of states with a concave function assigning an entropy to each state. However, as already explained, we need to let entropy to take values in $\extreal$ so that we can treat the heat bath as a thermostatic system and also take suprema of arbitrary sets of entropies. We thus make the following definition:

\begin{definition}
    A \define{thermostatic system} $(X, S)$ is a convex space $X$ together with a concave function $S \maps X \to \extreal$, where $\extreal$ has the convex structure given in \cref{ex:extended_reals}. We call $X$ the \define{state space}, call points of $X$ \define{states}, and call $S$ the \define{entropy} function.
\end{definition}

There are many examples of thermostatic systems coming from classical thermodynamics; here is a small sampling.

\begin{figure}
    \centering
    \includegraphics[width=0.5\textwidth]{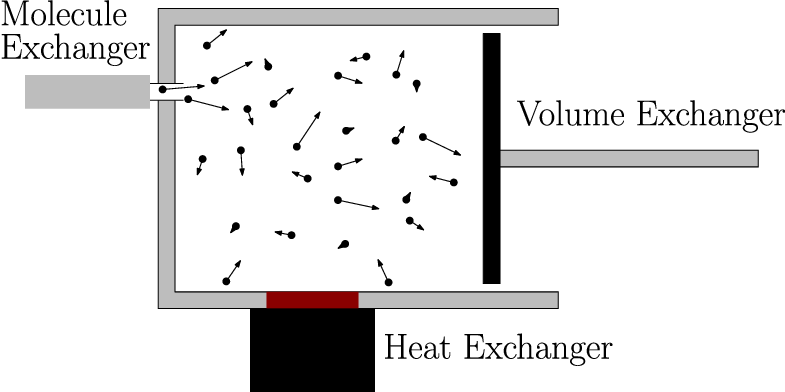}
    \caption{The Ideal Gas}
    \label{fig:ideal_gas}
\end{figure}

\begin{example}
\label{ex:ideal_gas}
    The ideal gas is a familiar thermostatic system with state space $X = \pos^3$, whose coordinates $(U, V, N)$ describe the energy, volume, and particle number of the gas. The entropy $\Sgas \maps X \to \extreal$ of the ideal gas is given by the Sackur--Tetrode equation \cite{SackurTetrode}. If we were to set up an experiment to study properties of the ideal gas, we would want to be able to change these three parameters. A theoretical setup for such an experiment is pictured in \cref{fig:ideal_gas}.
\end{example}

\begin{example}
\label{ex:tank}
    A closed tank of an idealized incompressible liquid can change neither its volume nor its number of particles; its state is solely determined by its total energy $U$. Thus, the state space for this system is $X= \pos$. The entropy is defined as $\Stank(U) = C\log(U)$ for $C \in \pos$ a constant. Classically, the temperature $T$ of a system is defined as
    \[ \frac{1}{T} = \pdv{S}{U} \]
    so we get
    \[ \frac{1}{T} = \frac{C}{U} \]
    and hence
    \[ CT = U. \]
    Therefore, we can identify $C$ with the heat capacity of the system. Each unit of increase in temperature leads to an increase in energy by $C$.
\end{example}

\begin{example}
\label{ex:heat_bath}
    The heat bath at constant temperature $T_0 \ne 0$ is an important thermostatic system that our formalism treats on an equal footing with others. It has state space $X = \real$ and entropy function $S(U) = U / T_0$. We think of the single coordinate $U$ not as the total energy of the heat bath (which is infinite), but rather the net energy transferred in or out of the heat bath. The infinite nature of the heat bath is expressed by the fact that $U$ can be arbitrarily negative: we can take out as much heat as we want. Moreover, taking the derivative of $S$ with respect to $U$, we find that
    \[ \pdv{S}{U} = \frac{1}{T_0} .\]
    Thus, the temperature is constant at $T_0$, no matter how much energy we put in or take out of the heat bath.
\end{example}

We can derive the heat bath as a certain limit of the tank, provided that we rescale properly. Fix a temperature $T$. The tank with heat capacity $C$ reaches temperature $T$ when it has energy $U = CT$. Now, consider for each $C$, the thermostatic system with state variable $\Delta U \in [-CT,\infty)$, and entropy function $S_C(\Delta U) = C\log(CT + \Delta U) - C\log(CT)$. For any fixed $\Delta U$, as $C \to \infty$, we have
\begin{align*}
    C\log(CT + \Delta U) - C\log(CT) &= C\log(\frac{CT + \Delta U}{CT}) \\
    &= \log(\qty(1 + \frac{1}{C}\frac{\Delta U}{T})^C) \\
    &\to \log(\exp(\frac{\Delta U}{T})) \\
    &= \frac{\Delta U}{T} 
\end{align*}
Thus, $S_C$ converges pointwise to $\Sbath$ at temperature $T$. This means that a tank with a high heat capacity behaves like a heat bath for small fluctuations around a given energy level.

There are also many important examples of thermostatic systems coming from statistical mechanics. Thus, our approach puts classical thermodynamics and statistical mechanics on an equal footing. In the following examples, and indeed throughout the paper, we use units where Boltzmann's constant $k$ is $1$. 

\begin{example}
\label{ex:Shannon}
    There is a thermostatic system whose state space is the set of probability distributions on $\{0,\dots, n\}$, namely the convex space $\Delta^n$ described in Example \ref{ex:simplex}. The natural choice of entropy function here is the \define{Shannon entropy}, given by
    \[ \Ssh(p) = -\sum_{i=0}^n p_i \log(p_i). \]
    This is well-known to be concave \cite[Thm.\ 2.73]{CoverThomas}.
\end{example}

\begin{example} 
\label{ex:measure_space}
    More generally, for any measure space $(X, \mu)$ the set of probability distributions $P(X, \mu)$ is a convex space, and this becomes a thermostatic system with entropy function $S \maps P(X, \mu) \to \extreal$ given by 
    \[   S(p) = -\int_X p(x) \log(p(x)) \, d\mu(x) .\]
\end{example}

\begin{example}
    We can also treat infinite-volume statistical mechanics with this framework. Let $\Omega = \set{-1,1}^{\integer^d}$. The weak topology on $\Omega$ is given by $\omega^{(n)} \to \omega$ iff $\omega^{(n)}_i \to \omega_i$ for all $i \in \integer^d$. Let $\Sigma$ be the Borel $\sigma$-algebra corresponding to this topology, and let $\mathcal{M}_{1,\theta}(\Omega)$ be the convex space of translation-invariant probability measures on this $\sigma$-algebra, where $\mu$ is translation-invariant if $\mu(U) = \mu(\set{\omega \circ \phi \st \omega \in U})$ for all translations $\phi \maps \integer^d \to \integer^d$.
    
    Then for any finite $\Lambda \ins \integer^d$, and any $\mu \in \mathcal{M}_{1,\theta}(\Omega)$, let $\mu_\Lambda$ be $\mu$ restricted to $\set{-1,1}^\Lambda$. We use this to define the entropy density of a translation-invariant measure $\mu$ as
    \[ s(\mu) = \limas_{n \to \infty} \frac{\Ssh(\mu_{B(n)})}{\abs{B(n)}} \]
    where $B(n)$ is the ball of radius $n$ in $\integer^d$. This is well-defined and concave \cite[Proposition 6.75]{FriedliVelenik}, so $(\Omega,s)$ is a thermostatic system.
\end{example}

\begin{example} 
\label{ex:density_matrices}
    For any Hilbert space $H$, let $X$ be the set of \define{density matrices} on $H$, i.e.\ nonnegative self-adjoint operators $\rho$ with $\mathrm{Tr}(\rho) = 1$. Then this becomes a thermostatic system with entropy function $S \maps X \to \extreal$ given by the \define{von Neumann entropy}
    \[  S_{\mathrm{vn}}(\rho) = -\mathrm{Tr}(\rho \log(\rho)) .\]
\end{example}

More generally still, concave entropy functions can be defined on the convex spaces of states in a large class of `generalized probabilistic theories', making them into thermostatic systems \cite{Barnum_et_al,Krumm_et_al,ShortWehner}. 

\begin{example}
\label{ex:generalized}
    Fix a convex space $X$. Define a \define{measurement} to be a convex-linear map $e \maps X \to \Delta^n$ for some $n$, where the $n$-simplex $\Delta^n$ is defined as in \cref{ex:simplex}. Thus, for each state $X$ the measurement gives a probability distribution on the set of \define{outcomes}, $\{0, \dots, n\}$. 
    
    Given any convex space $X$ equipped with a collection $E$ of measurements, we can define an entropy function $S_E \maps X \to [0,\infty)$ as follows:
    \[    S_E(x) = \inf_{e \in E} \Ssh(e(x)) \]
    Because the infimum of concave functions is concave, $S_E$ is concave. Thus, $(X,S_E)$ is a thermostatic system.
    
    Barnum \textit{et al.} \cite{Barnum_et_al} take this approach and do not impose any restriction on the collection $E$. Note however that `uninformative' measurements tend to drive down the entropy function $S_E$. For example, if $E$ includes the unique measurement with a single outcome, $f \maps X \to \Delta^0$, the entropy function $S_E$ is identically zero. 
    To prevent uninformative measurements from driving down the entropy, Short and Wehner take $E$ to be a collection of measurements that are `fine-grained' in a certain precise sense \cite{ShortWehner}. They argue that with this restriction, $S_E$ equals the usual Shannon entropy when $X = \Delta^n$ for some $n$, and the von Neumann entropy when $X$ is the set of density matrices on a finite-dimensional Hilbert space $H$. 
\end{example}

\section{Entropy maximization}
\label{sec:entropy_maximization}

It is well known that a system in thermodynamic equilibrium maximizes entropy subject to the constraints imposed on its states. The key insight behind our approach is that the constraints used in entropy maximization are typically \emph{parameterized}. For instance, in the system with two components that are constrained to have a fixed total energy, the total energy parameterizes this constraint. The formal structure that describes a parameterized collection of constraints is a convex relation $R \ins X \times Y$. This convex relation assigns to each $y \in Y$ a constrained set $\set{ x \in X \st (x, y) \in R}$.   In the example of a two-component system with a fixed total energy, $X = X_1 \times X_2$ is the convex set of states of the whole system, while $Y$ is the set of possible energies, and $(x,y) \in R$ when the total energy of both components equals $y$.

Recall that convex spaces and convex relations form a category $\convrelC$. We use this category to formalize this application of the maximum entropy principle by constructing a functor 
\[ \Ent \maps \convrelC \rightarrow \setC \]
sending any convex space $X$ to the set of all concave functions $S \maps X \to \extreal$. Then, given a concave function $S \maps X \to \extreal$ and a convex relation $R \ins X \times Y$, we define the function $R_*S \maps Y \to \extreal$ by
\[   R_*S(y) = \sup_{(x, y) \in R} S(x) .\]
In this way, we can ``coarse-grain'' the thermostatic system $(X, S)$ to make a system $(Y, R_*S)$. The entropy assigned to $y \in Y$ is the supremum of the entropies of all the states in $X$ ``compatible'' with $y$: that is, related to $y$ by the relation $R$.

\begin{lemma}
\label{thm:functor}
As defined above, $\Ent$ is a functor from $\convrelC$ to $\setC$.
\end{lemma}

\begin{proof}
    First we check well-definedness. Fix convex spaces $X$ and $Y$, a concave function $S \maps X \to \extreal$ and a convex relation $R \subseteq X \times Y$; we must show the function $R_*S$ defined above is concave. That is, we must show that for all $y, y' \in Y$ and $0 \le \lambda \le 1$ we have
    \[ R_*S(c_\lambda(y, y')) \geq c_\lambda(R_*S(y), R_*S(y')) .\]
    We consider several cases. Note that we need only consider $\lambda \in (0, 1)$; if $\lambda = 0$ or $\lambda = 1$ then the inequality is trivially true as an equation.
    \begin{enumerate}
        \item Suppose $R_*S(y), R_*S(y') \in \real$. Then for any $\epsilon > 0$, by definition of $R_*S$ as a supremum, we can choose $x$ and $x'$ such that $(x, y), (x', y') \in R$ and
        \begin{align*}
            S(x) &> R_*S(y) - \epsilon, \\
            S(x') &> R_*S(y') - \epsilon.
        \end{align*}
        Now fix $\lambda \in (0, 1)$. By the convexity of $R$, $(c_\lambda(x, x'), c_\lambda(y, y')) \in R$. It follows that
        \begin{align*}
            c_\lambda(R_*S(y), R_*S(y')) - \epsilon
            &= c_\lambda(R_*S(y)-\epsilon, R_*S(y') - \epsilon)
            \\&\leq c_\lambda(S(x), S(x'))
            \\&\leq S(c_\lambda(x, x'))
            \\&\leq R_*S(c_\lambda(y, y')).
        \end{align*}
        The second to last inequality is by concavity of $S$, and then the last inequality is by definition of $R_* S$. Letting $\epsilon \to 0$, we have our desired inequality.
        \item Suppose $R_*S(y) = \infty, R_*S(y') \in \real$. Then we can choose some $(x', y') \in R$ and choose $x \in X$ such that $(x,y) \in R$ with $S(x)$ positive and as large as we like. Now fix $\lambda \in (0, 1)$. Then we have
        \begin{align*}
            \lambda S(x) + (1-\lambda) S(x')
            &\leq S(c_\lambda(x,x'))
            \\&\leq R_*S(c_\lambda(y,y')).
        \end{align*}
        Since $R_*S(c_\lambda(y,y'))$ is bounded below by a quantity that is as large as we like, it is infinite. Thus the desired inequality holds.
        \item Suppose either $R_\ast S(y)$ or $R_\ast S(y')$ is $-\infty$. In this case the inequality to be proved is trivial since for $\lambda \in (0,1)$ we have $c_\lambda(R_\ast S(y),R_\ast S(y')) = -\infty$.
    \end{enumerate}
    Without loss of generality, all cases are equivalent to one of these three, so we have proved concavity.
    
    Next we check that $\Ent$ is a functor. We must show that given convex relations $R \subseteq X \times Y$, $R' \subseteq Y \times Z$ we have
    \[   (R' \circ R)_* = R'_* \circ R_* .\]
    The composite relation is defined so that if we fix $z \in Z$, we have
    \[ \{x \in X \mid (x, z) \in R' \circ R\} = \{x \in X \mid \exists y \in Y \; (x, y) \in R \textrm{ and } (y, z) \in R'\}. \]
    It follows that
    \begin{align*}
        (R'_* \circ R_*) (S)(z)
        &= \sup_{(y, z) \in R'} R_*(S)(y)
        \\&= \sup_{(y, z) \in R'} \left( \sup_{(x, y) \in R} S(x) \right)
        \\&= \sup_{(x, z) \in R' \circ R} S(x)
        \\&= (R' \circ R)_* (S)(z)
    \end{align*}
    showing that $\Ent$ preserves composition. Identity maps are clearly preserved.
\end{proof}

\begin{example}
    Consider a thermostatic system consisting of two tanks, with energy $U_1$ and $U_2$ respectively. The state space for this thermostatic system is $\pos^2$, and the entropy function is
    \[ S(U_1,U_2) = C_1 \log(U_1) + C_2 \log(U_2) \]
    where $C_1$ and $C_2$ are the heat capacities of the two systems, respectively.
    
    Now, consider the convex relation $R \ins \pos^2 \times \pos$ given by the equation
    \[ U_1 + U_2 = U.\]
    This relation lets us coarse-grain the thermostatic system with state space $\pos^2$ so that we only consider the total energy. If we push $S$ forward along $R$, we get
    \[ R_\ast S(U) = \sup_{U_1 + U_2 = U} S(U_1, U_2) \]
    The meaning of this is that the entropy of the coarse-grained state $U$ is the supremum of the entropies of the fine-grained system states $(U_1,U_2)$ compatible with $U$.
    
    This supremum is in fact achieved when
    \[ \pdv{U_1} C_1 \log(U_1) = \pdv{U_2} C_2 \log(U_2) \]
    so
    \[ \frac{C_1}{U_1} = \frac{C_2}{U_2}. \]
    Since $U_i/C_i$ is the temperature of tank $i$, this says that the temperatures of the two tanks are equal at equilibrium. If we substitute $U - U_1$ for $U_2$ we get
    \[  \frac{C_1}{U_1} = \frac{C_2}{U - U_1} \]
    and thus
    \[ U_1 = \frac{C_1}{C_1 + C_2} U , \qquad U_2 = \frac{C_2}{C_1 + C_2} U. \]
    This gives an explicit formula for $R_\ast S$ as a function of $U$:
    \begin{align*}
    R_\ast S(U) &= C_1 \log(\frac{C_1}{C_1 + C_2} U) + C_2 \log(\frac{C_2}{C_1 + C_2}U) \\
    &= (C_1 + C_2)\log(U) + K
    \end{align*}
    for some constant $K$ depending on $C_1$ and $C_2$. As maximization behavior does not change if we add a constant to entropy, this entropy function gives the same behavior as a tank of heat capacity $C_1 + C_2$, as expected.
\end{example}

In this example, we saw how entropy maximization could be used to compose two tanks. However, in order to construct this example, we had to use another general principle: the entropy of two independent systems is the sum of their individual entropies.  We would like our framework to incorporate this principle.  Our eventual goal is to be able to take multiple thermostatic systems, compose them with some constraints, and end up obtaining a single thermostatic system---in an automatic way, with no further decisions required. The mathematical constructs we will use for this are \emph{operads} and \emph{operad algebras}. However, we do not assume that the reader has prior familiarity with the theory of operads. Thus, the next section reviews operads, before developing the operad algebra of thermostatic systems. 

Before we move on to that, however, we give two examples that clarify the meanings of infinite and negative infinite entropy.

\begin{example} \label{ex:infty}
    Let $\Stank \maps \pos \to \extreal$ be the entropy of a closed tank of incompressible fluid as a function of its internal energy, given as in \cref{ex:tank} by $\Stank(U) = C \log(U)$. Consider the convex relation $R \subseteq \pos \times \set{\ast}$ given by allowing all elements of $\pos$ to be related to $\set{\ast}$. Pushing the tank's entropy forward along this relation, we obtain
    \[ R_\ast \Stank(\ast) = \sup_{U > 0} \Stank(U) = \infty.\]
    This illustrates the meaning of infinite entropy: when a thermostatic system can reach states of arbitrarily high entropy, its entropy in equilibrium is $\infty$.
\end{example}

\begin{example} \label{ex:-infty}
    In the spirit of the last example, consider a relation $R \subseteq \pos \times \real$ given by the graph of the inclusion $\pos \hookrightarrow \real$. Let $S$ be any entropy function on $\pos$. Then for any $x \le 0$,
    \[ R_\ast S(x) = \sup_{y > 0, y = x} S(y) = -\infty \]
    because the supremum of the empty set is $-\infty$. This illustrates the meaning of negative infinite entropy: an entropy of $-\infty$ represents an impossible state.
\end{example}

\section{The operad algebra of thermostatic systems}
\label{sec:operads}

In this section we construct an operad $\Op(\convrelC)$ where the operations are convex relations from a product of several convex spaces to a single convex spaces. These operations serve as ways to combine several thermostatic systems into a single such system. We formalize this fact by proving that the collection of all thermostatic systems forms an `algebra' of the operad $\Op(\convrelC)$. This algebra is called $\Op(\Ent)$, for two reasons. Intuitively, the principle whereby thermodynamic systems are combined is entropy maximization. Formally, $\Ent$ is the functor from the category $\convrelC$ to $\setC$ that assigns to any convex set $X$ the set of all entropy functions on $X$. In our main result, Theorem \ref{thm:operad_algebra}, we show that this functor defines an algebra $\Op(\Ent)$ of the operad $\Op(\convrelC)$.

Operads are a generalization of categories where the domain of a morphism is a family of objects, but the codomain is still required to be a single object. Operads originally arose in the study of iterated loop spaces \cite{May}, and continue to find use in homotopy theory and higher category theory. Recently, operads have also appeared in applied category theory \cite{NetMod, FoleyBreinerSubDusel, Spivak}. More detail on operads may be found in \cite{MarklShniderStasheff, SetOperads, Yau}. 

\begin{definition}
    An \define{operad} (also known as a symmetric multicategory) is a collection $\O_0$ of \define{types}, and for any types $X_1, \ldots, X_n, Y \in \O_0$ a collection of \define{operations} $\O(X_1, \ldots, X_n; Y)$ satisfying the following properties.
    \begin{itemize}
        \item Given operations
        \[f_1 \in \O(X_{1, 1}, \ldots, X_{1, m_1}; Y_1), \ldots, f_n \in \O(X_{n, 1}, \ldots, X_{n, m_n}; Y_n)\]
        \[g \in \O(Y_1, \ldots, Y_n; z)\]
        one can compose to construct a new operation
        \[g(f_1, \ldots, f_n) \in \O(X_{1, 1}, \ldots, X_{n, m_n}; z).\]
        \item For every type $X$ there is an identity operation $1_X \in \O(X; X)$.
        \item Composition must be associative and unital with respect to the identity operations.
        \item For every permutation $\sigma \in S_n$ there is a bijection $\O(X_1, \dots, X_n; Y) \to \O(X_{\sigma(1)}, \dots, X_{\sigma(n)}; Y)$ which must satisfy certain compatibility conditions \cite{Yau}.
    \end{itemize}
\end{definition}

\begin{figure}
    \centering
    \includegraphics[width=0.8\textwidth]{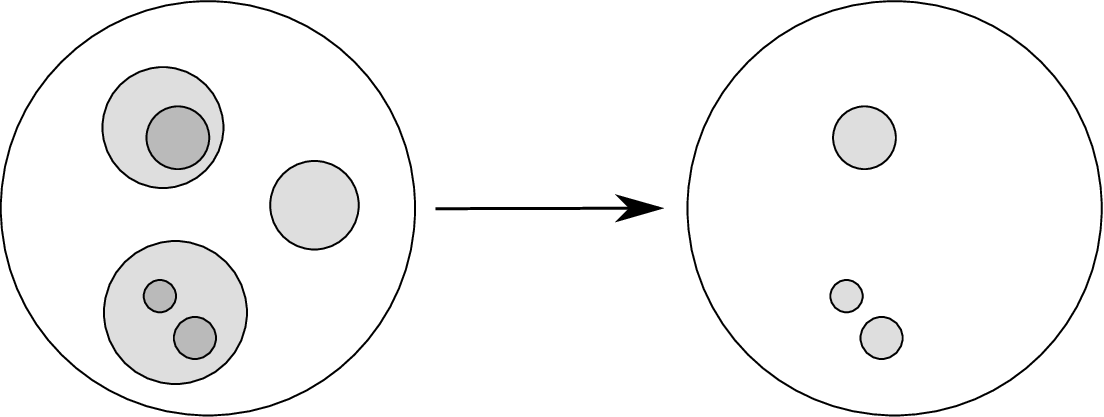}
    \caption{Composition in the little 2-disks operad.}
    \label{fig:little_2disks_composition}
\end{figure}

\begin{example}
    The \define{little 2-disks operad} is a famous operad which arises in topology  \cite{Sinha}. It has only one type, $X$, and an operation $X^n \to X$ is a labelled list of $n$ disjoint closed disks in the unit disk of $\real^2$. The composition of operations is simply composition of inclusions, and can be seen in \cref{fig:little_2disks_composition}. This operad does not play a role in our present framework, but one can draw similar looking pictures of operations in the operad for thermostatic systems.
\end{example}

Symmetric monoidal structures on categories are another formalism that allows one to discuss morphisms with multiple inputs, and to permute these inputs \cite{MacLane}, and in fact there is in fact a strong relationship between operads and symmetric monoidal categories. Every symmetric monoidal category has an underlying operad.

\begin{construction}
\label{con:smcoperad}
    Given a symmetric monoidal category $(\C, \otimes)$, there is an operad $\Op(\C)$ defined by
    \begin{align*}
        \Op(\C)_0 &= \C_0 \\
        \Op(\C)(X_1, \ldots, X_n; Y) &= \C(X_1 \otimes \cdots \otimes X_n, Y)
    \end{align*}
    The composition of the operad is constructed from the composition and monoidal product of the category, and similar for identities.
\end{construction}

\begin{example} \label{ex:operad_of_sets}
 Applying \cref{con:smcoperad} to the cartesian monoidal category $(\setC, \times, 1)$, we obtain an operad $\mathcal{S}$ of sets, where the types are sets, and the operations are multivariable functions: $\mathcal{S}(X_1, \ldots, X_n; Y) = \setC(X_1 \times \ldots \times X_n, Y)$.
\end{example}

\begin{example}
    In this example we put a symmetric monoidal structure on $\convrelC$ and then apply \cref{con:smcoperad} to get an operad whose types are convex spaces and whose operations are convex relations $R \subseteq (X_1 \times \cdots \times X_n) \times Y$.
    
    In \cref{def:product_convex_spaces} we explained the product of convex spaces. This is in fact the categorical product in $\convC$, the category of convex spaces and functions between them. Thus $(\convC, \times)$ is a symmetric monoidal category \cite{MacLane}. The category $\convrelC$ has the same objects as $\convC$, but more morphisms: convex relations rather than just convex maps. Like $\convC$, $\convrelC$ is a symmetric monoidal category where the tensor product of convex spaces is $X$ and $Y$ is $X \times Y$. However, this tensor product is no longer the categorical product in $\convrelC$, just as the usual cartesian product of sets is not the category product in $\relC$, the category of sets and relations. Thus we need to define the tensor product of morphisms in $\convrelC$ `by hand'. Luckily this is easy: the usual product of subsets defines a product of relations, and the product of two convex relations is again convex. We also need to endow $\convrelC$ with some natural isomorphisms: the associator
    \[   \alpha_{X,Y,Z} \maps (X \times Y) \times Z \to X \times (Y \times Z), \]
    the left unitor
    \[  \lambda_X \maps 1 \times X \to X \]
    where $1$ denotes a chosen singleton,
    and the right unitor
    \[   \rho_X \maps X \times 1 \to X , \]
    and the braiding
    \[   \beta_{X,Y} \maps X \times Y \to Y \times X .\]
    But these all maps are all the obvious ones---and they obey the necessary equations for a symmetric monoidal category because $(\convC, \times)$ is symmetric monoidal. Thus $(\convrelC, \times)$ is also symmetric monoidal, and we obtain an operad $\Op(\convrelC)$.
\end{example}

\begin{definition}
    A \define{map of operads} $F \maps \O \to \P$ consists of a map of types \[X \in \O_0 \mapsto F(X) \in \P_0\] and for every $X_1, \cdots, X_n, Y \in \O_0$, a map of operations \[f \in \O(X_1, \ldots, X_n; Y) \mapsto F(f) \in \P(F(X_1), \ldots, F(X_n); F(Y))\]
    This map of operations must commute with composition and identities, i.e.\ \[F(g(f_1, \ldots, f_n)) = F(g)(F(f_1), \ldots, F(f_n))\] and \[F(1_X) = 1_{F(X)}.\] Let $\oprdC$ denote the category of operads and their maps.
\end{definition}

\begin{definition}
    For an operad $\O$, an \define{$\O$-algebra} is a map of operads from $\O$ to $\Op(\setC)$.
\end{definition}

We can construct maps of operads, and indeed operad algebras, from `lax symmetric monoidal functors': see Mac Lane's text \cite{MacLane} for these.

\begin{construction}
\label{construction:operad_algebra}
    Given a lax symmetric monoidal functor $(G, \epsilon) \maps (\C, \otimes_\C) \to (\D, \otimes_\D)$, we can construct a map of operads $\Op(G) \maps \Op(\C) \to \Op(\D)$ in the following manner. For an object $X$ of $\C$, define $\Op(G)(X) = G(X)$. For a morphism $f \in \Op(C)(X_1, \ldots, X_n; Y)$ (i.e.\ $f \maps X_1 \otimes_\C \cdots \otimes_\C X_1 \to Y$), define
    \[\Op(G)(f) = G(X_1) \otimes_\D \cdots \otimes_\D G(X_n) \xrightarrow{\epsilon_{X_1, \ldots, X_n}} G(X_1 \otimes_\C \cdots \otimes_C X_n) \xrightarrow{G(f)} G(Y).\]
    In this way, $\Op$ defines a functor $\SMC_\ell \to \oprdC$.
    In the case that $G$ is a lax symmetric monoidal functor to $\setC$, then $\Op(G)$ is an $\Op(\C)$-algebra.
\end{construction}

Using this construction, we can prove that the functor $\Ent \maps \convrelC \to \setC$ from \cref{thm:functor} defines an operad algebra of the operad $\Op(\convrelC)$ by showing that $\Ent$ is a lax symmetric monoidal functor.

To do this, we need to equip the functor $\Ent$ with a `laxator' 
\[ \epsilon_{X_1, X_2} \maps \Ent(X_1) \times \Ent(X_2) \to \Ent(X_1 \times X_2)  \]
and a map $\epsilon_0 \maps 1 \to \Ent(1)$. 
Given functions $S_1 \in \Ent(X_1)$ and $S_2 \in \Ent(S_2)$ we can define an element $S_1+S_2\in \Ent(S_1 \times S_2)$ as follows:
\[ (S_1+S_2)(x_1,x_2) = S_1(x_1) + S_2(x_2) \]
where addition in $\extreal$ is defined as usual for real numbers, but we set
 \[ 
    \begin{array}{ccccrcl}
        x + \infty &=& \infty + x &=& \infty &\quad & \text{for $x \in \real$} \cup \set{\infty}\\ 
        x + (-\infty) &=& -\infty + x &=& -\infty &\quad& \text{for $x \in \real$} \cup \set{-\infty}\\
        \infty + (-\infty) &=& -\infty + \infty &=& 
        -\infty.
    \end{array}
    \]
Thus, as in the convex structure, negative infinity ``dominates'' positive infinity. The map $\epsilon_{X_1,X_2}$ is defined to map $(S_1,S_2)$ to $S_1+S_2$. The map $\epsilon_0 \maps 1 \to \Ent(1)$ simply picks out the constant $0$ function on the singleton.

\begin{lemma}
\label{thm:lax}
    The natural transformation $\epsilon_{X_1, X_2} \maps \Ent(X_1) \times \Ent(X_2) \to \Ent(X_1 \times X_2)$ and the map $\epsilon_0 \maps 1 \to \Ent(1)$ define a lax symmetric monoidal structure on the functor $\Ent \maps \convrelC \to \setC$.
\end{lemma}

\begin{proof}
    First we show that $\epsilon$ is natural. Let $R_1 \maps X_1 \to Y_1$ and $R_2 \maps X_2 \to Y_2$ be convex relations, and let $S_1 \in \Ent(X_1)$ and $S_2 \in \Ent(X_2)$ be concave functions. We need to show the following square commutes.
    \[
    \begin{tikzcd}
        \Ent(X_1) \times \Ent(X_2) 
        \arrow[d, swap, "R_{1*} \times R_{2*}"]
        \arrow[r, "\epsilon_{X_1, X_2}"]
        &
        \Ent(X_1 \times X_2)
        \arrow[d, "(R_1 \times R_2)_*"]
        \\
        \Ent(Y_1) \times \Ent(Y_2)
        \arrow[r, swap, "\epsilon_{Y_1, Y_2}"]
        &
        \Ent(Y_1 \times Y_2)
    \end{tikzcd}
    \]
    
\begin{figure}
    \centering
    \includegraphics[width=0.4\textwidth]{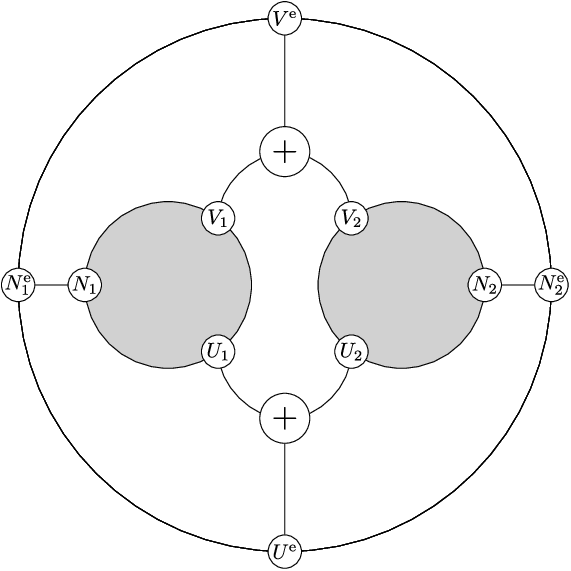}
    \caption{The operation of pressure and temperature equalization.}
    \label{fig:two_gases_equalize_pressure}
\end{figure}    
    We compute:
    \begin{align*}
        (R_1 \times R_2)_* \circ \epsilon_{X_1, X_2} (S_1, S_2) (y_1, y_2)
        &= (R_1 \times R_2)_* (S_1+S_2) (y_1, y_2)
        \\&= \sup_{(x_1, x_2)(R_1 \times R_2)(y_1, y_2)} (S_1+S_2) (x_1, x_2)
        \\&= \sup_{(x_1,y_1) \in R_1, (x_2,y_2) \in R_2} S_1(x_1) + S_2(x_2)
        \\&= \sup_{(x_1,y_1) \in R_1} S_1(x_1) + \sup_{(x_2,y_2) \in R_2} S_2(x_2)
        \\&= R_{1*}S_1(y_1) + R_{2*}S_2(y_2)
        \\&= (R_{1*}S_1 + R_{2*}S_2) (y_1, y_2)
        \\&= \epsilon_{Y_1, Y_2} (R_{1*}S_1, R_{2*}S_2) (y_1, y_2)
        \\&= \epsilon_{Y_1, Y_2} \circ (R_{1*} \times R_{2*}) (S_1, S_2) (y_1, y_2).
    \end{align*}

    To show that $(\Ent,\epsilon,\epsilon_0)$ is a lax symmetric monoidal functor from $\convrelC$ to $\setC$ we need to check some equations, which are explained in Mac Lane's text \cite{MacLane}. For this we need to use the fact that addition makes $\extreal$ into a commutative monoid with $0$ as its identity element. To see this, note that if $X$ is any commutative monoid, there is a unique commutative monoid structure on $X \sqcup \{\ast\}$ extending that on $X$ such that $x + \ast = \ast$ for all $x \in X$. Using this once, we get a commutative structure on $(-\infty,\infty] = \real \sqcup \{\infty\}$, and using it again, we get the desired commutative monoid structure on $\extreal = (-\infty,\infty] \sqcup \{-\infty\}$. 
    
    Let $S_i \in \Ent(X_i)$, for $i=1,2,3$. First we need to check that $\epsilon$ obeys the hexagon identity relating it to the associator $\alpha_{X_1,X_2,X_3}$ in $\convrelC$ and the associator $\alpha_{\Ent(X_1),\Ent(X_2),\Ent(X_3)}$ in $\setC$:
\begin{align*}
        (\alpha_{X_1, X_2, X_3})_* \circ \epsilon_{X_1 \times X_2, X_3} \circ & [\epsilon_{X_1, X_2} \times 1_{\Ent X_3}] ((S_1, S_2), S_3)
        = (\alpha_{X_1, X_2, X_3})_* \circ \epsilon_{X_1 \times X_2, X_3} (S_1+S_2, S_3) \\
  &= (\alpha_{X_1, X_2, X_3})_* ((S_1+S_2)+S_3) \\
&= S_1+(S_2+S_3) \\
&= \epsilon_{X_1, X_2 \times X_3} (S_1, S_2+S_3) \\
&=  \epsilon_{X_1, X_2 \times X_3} \circ [1_{\Ent(X_1)} \times \epsilon_{X_2, X_3}] (S_1, (S_2, S_3)) \\
&= \epsilon_{X_1, X_2 \times X_3} \circ [1_{\Ent(X_1)} \times \epsilon_{X_2, X_3}] \circ \alpha_{\Ent(X_1), \Ent(X_2), \Ent(X_3)}((S_1, S_2), S_3).
\end{align*}
    
Next we need to check that $\epsilon$ obeys the triangle equation relating it to the left unitor $\lambda_X$ in $\convrelC$ and the left unitor $\lambda_{\Ent(X)} \maps 1 \times \Ent(X) \to \Ent(X)$ in $\setC$:

    \begin{align*}
        (\lambda_X)_* \circ \epsilon_{1, X} \circ [\epsilon_0 \times 1_{\Ent(X)}] (\star, S)
        &= (\lambda_X)_* \circ \epsilon_{1, X} (0, S)
        \\&= (\lambda_X)_* (0+S)
        \\&= S
        \\&= \lambda_{\Ent(X)}(\star, S)
    \end{align*}
    where $\star$ is the unique element of $1$. The triangle equation for right unitor works analogously.
    This proves that $(F,\epsilon,\epsilon_0)$ is a lax monoidal functor. 
    
    Finally, to show that this functor is lax symmetric monoidal, we need to check that $\epsilon_{X_1,X_2}$ is compatible with the braiding $\beta_{X_1,X_2}$ in $\convrelC$ and the braiding \[\beta_{\Ent(X_1),\Ent(X_2)} \maps \Ent(X_1) \times \Ent(X_2) \to \Ent(X_2) \times \Ent(X_1)\] in $\setC$:
    \begin{align*}
        (\beta_{X_1, X_2})_* \circ \epsilon_{X_1, X_2} (S_1, S_2)
        &= (\beta_{X_1, X_2})_* (S_1+S_2)
        \\&= S_2+S_1
        \\&= \epsilon_{X_2, X_1} (S_2, S_1)
        \\&= \epsilon_{X_2, X_1} \circ \beta_{\Ent(X_1), \Ent(X_2)}(S_1, S_2). \qquad \qquad \qquad \qedhere
    \end{align*}
\end{proof}

With the help of this lemma, we can now prove our main result.

\begin{theorem}
\label{thm:operad_algebra}  
Thermostatic systems form an operad algebra of the operad of convex relations. That is, the lax symmetric monoidal functor $\Ent \maps \convrelC \to \setC$ defines an operad algebra $\Op(\Ent)$ of the operad $\Op(\convrelC)$.
\end{theorem}

\begin{proof}
To apply Construction \ref{construction:operad_algebra} we only need that $\Ent$ is a lax symmetric monoidal functor, which was shown in Lemmas \ref{thm:functor} and \ref{thm:lax}.
\end{proof}

To understand the significance of this result it helps to consider many examples.

\begin{figure}
    \centering
    \includegraphics[width=0.8\textwidth]{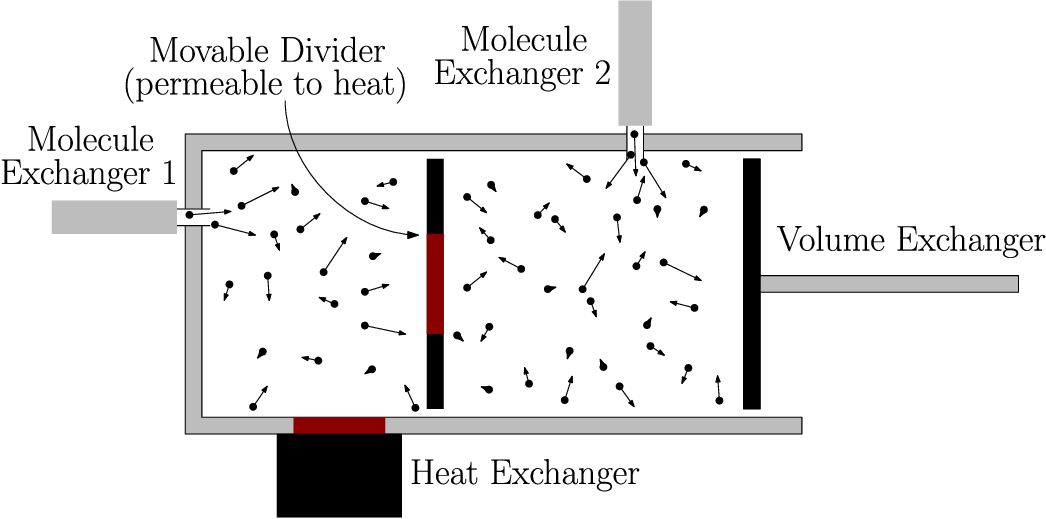}
    \caption{The physical setup for pressure and temperature equalization.}
    \label{fig:two_ideal_gases}
\end{figure}

\begin{example}
    In \cref{fig:two_gases_equalize_pressure}, we see a depiction of an operation 
    \[  R \in \Op(\convrelC)(\pos^3, \pos^3; \pos^4) \] 
    which takes two systems each having a volume, internal energy and particle number and composes them. This operation imposes a constraint on the total volume and total internal energy, while imposing no constraint on the particle numbers. Physically, this operation can be implemented with chambers of gas as in \cref{fig:two_ideal_gases}.
    
    When interpreted via the operad algebra $\Op(\Ent)$, this operation becomes one that takes in concave entropy functions $S_1 \maps \pos^3 \to \extreal$ and $S_2 \maps \pos^3 \to \extreal$ and constructs a new entropy function 
    \[ S = \Ent(R)(S_1, S_2) \maps \pos^4 \to \extreal. 
    \] We think of the two input entropy functions $S_1, S_2$ as ``filling in'' the disks in the middle of the diagram in \cref{fig:two_gases_equalize_pressure}, and $S$ as the entropy function of the large disk.

    The convex relation $R$ is given by
    \[ \begin{array}{clcl}
        U^e &= U_1 + U_2 \\
        V^e &= V_1 + V_2 \\
        N_1^e &= N_1 \\
        N_2^e &= N_2.
    \end{array}
    \]
    By the definition of $\Ent$, the entropy function for the whole system, $S(U^e, V^e, N_1^e, N_2^e)$, is the supremum of $S_1(U_1, V_1, N_1) + S_2(U_2, V_2, N_2)$ subject to the above constraints. These constraints can be rephrased as $V_2 = V^e - V_1$ and $U_2 = U^e - U_1$. Thus, we can formulate this problem as the problem of finding
    \[  S(U^e,V^e,N_1^e,N_2^3) = \sup_{U_1 \in [0,U^e], V_1 \in [0,V^e]} S_1(U_1, V_1, N_1^e) + S_2(U^e - U_1, V^e - V_1, N_2^e)\]
    Assuming that $S_1$ and $S_2$ are differentiable, and taking partial derivatives with respect to $U_1$ and $V_1$, we see that at any maximizing state not at the endpoints of the intervals above,
    \[\pdv{U_1} S_1(U_1, V_1, N_1^e) = \pdv{U_2} S_2(U_2, V_2, N_2^e) \]
    and 
    \[\pdv{V_1} S_1(U_1, V_1, N_1^e) = \pdv{V_2} S_2(U_2, V_2, N_2^e) \]
    Thus, the temperature and pressure have both equilibriated.
\end{example}

\begin{example}
\label{ex:legendre}
    In this example, we model the thermal connection of a gas to a heat bath held at a constant temperature $T$. The gas has state space $\pos^3$ containing states of the form $(U_{\mathrm{gas}}, V_{\mathrm{gas}}, N_{\mathrm{gas}})$, while the heat bath has state space $\real$ containing states $U_{\mathrm{bath}}$ as in \cref{ex:heat_bath}. We shall compose these to obtain a thermostatic system having state space $\pos^2$, where we can vary the volume and particle number but the energy is determined by the entropy maximization principle. 
    
    Thus, the process of composition is the operation $R \in \Op(\convrelC)(\pos^3,\real;\pos^2)$ given by the relation $R \ins (\pos^3 \times \real) \times \pos^2$ defined by the equations
    \begin{align*}
        U_{\mathrm{gas}} + U_{\mathrm{bath}} &= 0 \\
        V_{\mathrm{gas}} &= V \\
        N_{\mathrm{gas}} &= N
    \end{align*}
    where $(V,N)$ are coordinates on $\pos^2$. Given the entropy function of the gas, $S_{\mathrm{gas}} \maps \pos^3 \to \extreal$, and the entropy function of the heat bath, $S_{\mathrm{bath}} \maps \real \to \extreal$, the entropy of the composed system is
    \[   S = \Ent(R)(S_{\mathrm{gas}}, S_{\mathrm{bath}}) \maps \pos^2 \to \extreal.\]
    Since the entropy of the bath is
    \[ S_{\mathrm{bath}}(U_{\mathrm{bath}}) = U_{\mathrm{bath}}/T ,\]
    the entropy $S(V,N)$ is the supremum of
    \[S_{\mathrm{gas}}(U_\mathrm{gas},V,N) + \frac{U_\mathrm{bath}}{T}\]
    subject to the constraint $U_{\mathrm{gas}} + U_{\mathrm{bath}} = 0$. It follows that
    \[S(V,N) = \sup_{U_\mathrm{gas} \ge 0} \left( S_{\mathrm{gas}}(U_\mathrm{gas},V,N) - \frac{U_\mathrm{gas}}{T} \right).\]
    Viewed as a function of $V,N$ and $1/T$, this is a Legendre transform of $S_{\mathrm{gas}}$. Thus, composition of thermostatic systems can have the effect of taking a Legendre transform. 
\end{example}

\begin{example} \label{ex:canonical_ensemble}
    We can also connect a statistical mechanical system to a heat bath. Let $\Delta^n$ be the convex space of probability distributions on the finite set $\{0,\dots,n\}$. This becomes a thermostatic system with its Shannon entropy $\Ssh \maps \Delta^n \to \extreal$, defined as in \cref{ex:Shannon}. Let $\real$ be the space of states of a heat bath, made into a thermostatic system with entropy function $\Sbath(U) = U/T$ as in \cref{ex:heat_bath}.
    
    Let $H \maps \{0,\dots,n\} \to \real$ be any function, which we think of as a Hamiltonian. We write $H_i$ for the value of this function at $i \in \{0,\dots,n\}$. We can define a relation $R \ins (\Delta^n \times \real) \times 1$ by saying that $(p,U)$ is related to the one element $\ast \in 1$ iff
    \[ U + \sum_{i=0}^n H_i p_i = 0 .\]
    That is, the expected energy for our statistical mechanical system equals the energy taken from the heat bath. 
    
    A relation with $1$ is really just the same thing as a subspace of the domain; it is an `unparameterized' constraint. Thus, the entropy
    \[    S = \Ent(R)(\Ssh,\Sbath) \maps \ast \to \extreal \]
    amounts a single extended real number: the maximum possible entropy of the statistical mechanical system combined with the heat bath at temperature $T$. Note that the constraint $U + \sum_{i=0}^n H_i p_i = 0$ implies
    \[ \Sbath(U) = U/T = -\beta \sum_{i=0}^n H_i p_i \]
    where $\beta = 1/T$. (Recall that we use units where Boltzmann's constant is 1.)  Thus, by the definition of $\Ent$,
    \[ S = \sup_{p \in \Delta^n} \left(\Ssh(p) - \beta \sum_{i=0}^n H_i p_i \right).\]
    As well known, the supremum is obtained when $p$ is the famous Boltzmann distribution
    \[p_i = \frac{\exp(-\beta H_i)}{\sum_{i=0}^n \exp(-\beta H_i)} .\]
    Thus, the Boltzmann distribution can be obtained by connecting a statistical mechanical system to a heat bath. The same idea applies to statistical mechanical systems where states are either probability distributions on general measure spaces (\cref{ex:measure_space}) or density matrices (\cref{ex:density_matrices}).
\end{example}

In \cref{ex:canonical_ensemble} we obtained the Boltzmann distribution, or `canonical ensemble', by connecting a statistical mechanical system to a heat bath. We can also obtain the grand canonical ensemble and microcanonical ensemble using our formalism.

\begin{example} \label{ex:grand_canonical_ensemble}
    The grand canonical ensemble is obtained by coupling a statistical mechanical system to both a heat bath and a `particle bath', which is a thermostatic system mathematically isomorphic to a heat bath, but with a different physical interpretation. 
    
    Thus, we start with the thermostatic system $\Ssh \maps \Delta^n \to \extreal$ as in  \cref{ex:canonical_ensemble}, but now we choose two functions $H,M \maps \set{0,\ldots,n} \to \real$, one sending each state to its energy, and the other sending each state to its number of particles. We then introduce two other thermostatic systems: as before, a heat bath 
    \[  \begin{array}{rcl}
    S_{\mathrm{bath},1} \maps \real &\to& \extreal \\  
    U &\mapsto& \beta U 
    \end{array} \]
    where $\beta \in \real$ is the inverse temperature, but now also a particle bath 
    \[  \begin{array}{rcl}
    S_{\mathrm{bath},2} \maps \real &\to& \extreal \\  
    N &\mapsto& \beta \mu N 
    \end{array} \]
    where $\mu \in \real$ is the chemical potential.
    
    We couple these three systems using a relation $R \subseteq (X \times \real \times \real) \times 1$ for which $(p,U,V)$ is related to $\ast \in 1$ iff
    \[ U + \sum_{i=0}^n H_i p_i = 0 \]
    and
    \[ N + \sum_{i=0}^n M_i p_i = 0 .\]
    Following reasoning like that of \cref{ex:canonical_ensemble}, the entropy
    \[    S_{\mathrm{grand}} = \Ent(R)(\Ssh,S_{\rm{bath},1}, S_{\rm{bath},2}) \maps \ast \to \extreal \]
    amounts to a single extended real number
    \[ S_{\mathrm{grand}} = \sup_{p \in X} \left(\Ssh(p) - \beta \sum_{i=0}^n H_i p_i - \beta \mu \sum_{i=0}^n M_i p_i \right).\]
    As well known, the supremum is obtained when $p$ is the grand canonical ensemble. 
\end{example}

\begin{example} \label{ex:microcanonical_ensemble}
    The microcanonical ensemble is a probability distribution that represents a system at a fixed energy $U$, thermally isolated from its environment.
    To derive the microcanonical ensemble from our formalism, start with the thermostatic system $\Ssh \maps \Delta^n \to \extreal$ and a Hamiltonian $H \maps \{0,\dots,n\} \to \real$. Define a relation $R \subseteq \Delta^n \times \real$ by saying that $p$ is related to $U$ iff $H_i = U$ for all $i \in \{0,\dots, n\}$ with $p_i > 0$.
    
    The entropy of the microcanonical ensemble is defined to be
    \[   S_{\textrm{micro}} = \Ent(R)(\Ssh) \maps \real \to \extreal .\]
    It follows that $S_{\textrm{micro}}(U)$ is the supremum of $\Ssh(p)$ over probability distributions having $H_i = U$ for all $i$ with $p_i > 0$. This supremum is attained by the uniform distribution over the set of $i$ having energy $U$. Thus, $S_{\textrm{micro}}(U) = \log n$ if there are $n$ choices of $i$ with energy $U$. If no such choices of $i$ exist, then  $S_{\textrm{micro}}(U) = -\infty$. This is another example of the point made in \cref{ex:-infty}: an entropy of $-\infty$ represents an impossible state.
\end{example}

\section{Conclusion}

Having shown that convex spaces equipped with concave entropy functions form a natural context for studying thermostatic systems and the operations of composing such systems, one obvious direction for further research involves the Legendre transform. This transform is essential for deriving the multitude of `thermodynamic potentials' used in thermodynamics, of which the most famous are Gibbs free energy, Helmholtz free energy and enthalpy \cite{GaglianiScotti, PointErlicher, Willerton}. As shown in Example \ref{ex:legendre}, the Legendre transform can be implemented in our framework by attaching a thermodynamic system to a `bath' system, either of heat, or pressure, or some other quantity. This is satisfying because it gives a new physical interpretation of the Legendre transform.  However, there is much left to do to understand how the Legendre transform is connected to our framework or some extension of it.

\printbibliography

\end{document}